\documentclass[sigconf]{acmart}

\usepackage{changes}
\usepackage{multirow}
\usepackage{tcolorbox}
\usepackage{enumitem}
\usepackage{subfigure} 
\usepackage{subcaption}
\usepackage{changes}
\usepackage{ulem}
\usepackage{array}
\usepackage{xcolor}
\usepackage{soul}

\AtBeginDocument{%
  }

\setcopyright{acmlicensed}
\copyrightyear{2025}
\acmYear{2025}
\acmDOI{XXXXXXX.XXXXXXX}

\acmConference[WSDM]{The 18th ACM International Conference on Web Search and Data Mining}{March 10--14,
  2025}{Hannover, Germany}
\acmISBN{978-1-4503-XXXX-X/18/06}

\begin{document}

\title{Explainable CTR Prediction via LLM Reasoning}




\author{Xiaohan Yu}
\affiliation{%
  \institution{Huawei Cloud BU}
  \city{Beijing}
  \country{China}}
  \email{yuxiaohan5@huawei.com}

\author{Li Zhang}
\affiliation{%
 \institution{Institute of Finance Technology, UCL}
 \institution{Civil, Environmental and Geomatic Engineering, UCL}
 \country{United Kingdom}}
 \email{ucesl07@ucl.ac.uk}

\author{Chong Chen}
\authornote{Corresponding authors}
\affiliation{%
  \institution{Huawei Cloud BU}
  \city{Beijing}
  \country{China}}
  \email{chenchong55@huawei.com}

\renewcommand{\shortauthors}{Xiaohan yu, Li Zhang, Chong Chen}

\begin{abstract}
Recommendation Systems have become integral to modern user experiences, but lack transparency in their decision-making processes. Existing explainable recommendation methods are hindered by reliance on a post-hoc paradigm, wherein explanation generators are trained independently of the underlying recommender models. This paradigm necessitates substantial human effort in data construction and raises concerns about explanation reliability.  In this paper, we present ExpCTR, a novel framework that integrates large language model based explanation generation directly into the CTR prediction process. Inspired by recent advances in reinforcement learning, we employ two carefully designed reward mechanisms, LC alignment, which ensures explanations reflect user intentions, and IC alignment, which maintains consistency with traditional ID-based CTR models. Our approach incorporates an efficient training paradigm with LoRA and a three-stage iterative process. ExpCTR circumvents the need for extensive explanation datasets while fostering synergy between CTR prediction and explanation generation. Experimental results demonstrate that ExpCTR significantly enhances both recommendation accuracy and interpretability across three real-world datasets.

\end{abstract}




\maketitle
\section{Introduction}

Recommendation Systems (RS) have become a cornerstone of modern user experiences, empowering users to discover relevant and personalized items or contents  \cite{jannach2010recommender}. Collaborative methods \cite{fm, mnih2007probabilistic, he2017neural} have been dominant in this field, leveraging user-item interaction data for future predictions. While these methods, ranging from simple collaborative approaches to deep neural networks, have demonstrated remarkable efficacy in predicting user engagement, particularly in tasks such as click-through rate (CTR) prediction, they often operate as "black boxes", offering recommendations without explaining the underlying rationale \cite{er_survey}. 
The imperative for transparency and accountability has given rise to the burgeoning of explainable recommendation, which moves beyond mere suggestions by providing justifications. Such explanations provide numerous benefits: building user trust and satisfaction, enhancing persuasiveness, and enabling effective debugging and refinement \cite{tintarev2007explanations}.
Currently, the prevailing approach to explainable recommendation relies on a post-hoc paradigm, where explanations are generated independently of the recommendation model after its predictions are made. These methods necessitate substantial human effort to curate external training datasets through customer review processing or handcrafted rules to produce human-readable explanations.

\begin{figure}
    \centering
    \includegraphics[width=1.0\linewidth]{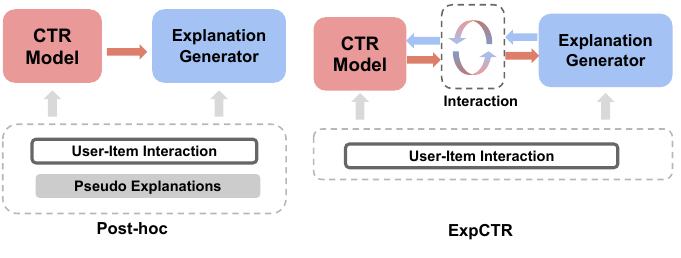}
    \caption{Comparison of current post-hoc paradigm methods and ours.}
    \label{fig:motivation}
\end{figure}

Recently, Large Language Models (LLMs) have emerged as a powerful tool in natural language processing, demonstrating exceptional reasoning capabilities. Their potential to generate human-readable explanations for complex tasks is particularly promising for explainable recommendation. 
Studies such as PETER \cite{peter} and RecExplainer \cite{lei2023recexplainer} have explored integrating item and user latent representations into pre-trained language models, harnessing collaborative information to enhance explanation generation. Other researchers probe the innate reasoning capabilities of LLMs for recommendation tasks using in-context learning techniques \cite{preliminary}. Chat-Rec \cite{chatrec} has showcased the potential of LLMs for improving explainability in multi-round conversational contexts.
Despite these promising developments, these approaches still rely on post-hoc explanations. They either over-rely on enhancing existing methods by substituting traditional language models with transformer-based LLMs or utilize basic zero-shot generation capabilities. Consequently, research on explainable recommendation with LLMs remains in its infancy. As illustrated in Figure \ref{fig:motivation}, several critical challenges persist:
\begin{itemize}[leftmargin=*]
    \item \textbf{Resource intensity.} Developing high-quality training datasets for explanation generators is resource-intensive, demanding substantial human effort. While customer reviews present a potential source of pseudo-explanations, they necessitate meticulous curation, extraction, and reformulation to yield training samples. Alternatively, methods like Chat-Rec necessitate extensive human involvement through interactive dialogues.

    \item \textbf{Explanation quality unreliability:} The post-hoc paradigm introduces potential discrepancies between the generated explanations and the underlying operations of recommender systems.
    Current methodologies typically employ a unidirectional information flow, where latent representations or prediction results are passed from the recommender model to a separate explanation generator \ref{fig:motivation}.
    This unidirectional process lacks mechanisms for quality assessment or feedback from the generated explanations to the existing recommender system.
    Consequently, there is no assurance that the produced explanations accurately reflect the recommender's internal decision-making process.

\end{itemize}

In light of the aforementioned challenges, we propose ExpCTR, a novel approach that aims to operate in a data-free manner, while fostering synergy between CTR prediction and LLM-based explanation generation. Our method seamlessly integrates LLM-driven explanation generation with the CTR prediction process. Drawing inspiration from recent advancements in reinforcement learning \cite{instructgpt}, we employ real-world feedback signals to refine the LLM's reasoning capabilities to better align with the objectives of CTR prediction.

ExpCTR involves a carefully crafted prompt template, tailored to fully elicit the LLM's reasoning capabilities through a chain-of-thought prompting strategy. Subsequently, we utilize a proximal policy optimization (PPO) algorithm that incorporates two distinct reward mechanisms: (1) LC alignment reward, which ensures that the produced explanations accurately reflect user intentions and preferences, as assessed by an LLM-based CTR predictor. (2) IC alignment reward, which treats the explanations as a textual input feature for a traditional ID-based CTR model, ensuring that the explanations are consistent with the model's internal mechanisms and predicted outcomes. These two rewards collectively incentivize the LLMs to generate explanations that are both human-centric and recommender-aligned.
To accommodate the reward designs, we devise a specific training paradigm that leverages LoRA for LLM lightweight fine-tuning. This paradigm is based on a three-stage iterative process, consisting of aligning with user interactions with LC alignment reward, training a CTR model with textual features, and aligning with the recommender system's internal mechanisms with IC alignment reward. These stages are iteratively repeated to progressively improve the ExpCTR's performance.
Our approach effectively circumvents the need for extensive explanation data construction and fosters collaboration between LLM-driven explainability and accurate CTR prediction. By deepening the understanding of user preferences and the recommendation mechanism, ExpCTR shows the potential to significantly enhance both interpretability and recommendation effectiveness.
Our key contributions can be summarized as follows:
\begin{itemize}[leftmargin=*]
    \item We introduce ExpCTR, an innovative framework that enhances the reasoning capabilities of LLMs to generate precise explanations that are closely aligned with CTR models. This approach simultaneously improves CTR prediction performance and RS interpretability. To the best of our knowledge, this represents the first attempt to leverage LLMs for this dual purpose without dependence on extensive data resources.
    \item We develop a reinforcement learning based approach to efficiently fine-tune LLMs using LoRA. Our approach integrates two meticulously designed reward mechanisms within a tailored three-stage training paradigm.
    \item We conduct a comparative analysis of ExpCTR against several state-of-the-art CTR prediction methods and evaluate the quality of the generated explanations, demonstrating the effectiveness of our method.
\end{itemize}

\section{Related Work}
Explainable recommendation (ER) extends traditional recommendation systems by addressing the "why" behind suggested items. ER provides not only item recommendations but also justifications clarifying the rationale for those suggestions \cite{zhang2020explainable}. Current methods can be broadly classified into two categories, model-intrinsic and post-hoc. Model-intrinsic methods aim for inherent explainability by leveraging interpretable algorithms \cite{er_survey}.
Conversely, post-hoc approaches leverage black-box models for recommendation, followed by a separate explanation model that deciphers the reasoning behind the recommendations. 
The rise of deep neural networks has propelled post-hoc methods to the forefront, transforming explainable recommendation into a natural language generation task.
Early works rely on pre-defined templates or association rules \cite{wang2018explainable, gao2019explainable}. Later advancements adopt Recurrent Neural Networks (RNNs) and Long Short-Term Memory (LSTM) architectures for generating textual explanations \cite{li2017neural, zhang2023recommendation}. With the advent of the Transformer architecture, researchers have explored their potential for explanation generation \cite{peter}. \cite{yang2024fine} incorporates reinforcement learning techniques to address potential issues like hallucinations. 
Despite these advancements, these approaches rely on generators trained independently with carefully curated explanation datasets. Given the scarcity of user-item-explanation triplets in real-world RS, substantial efforts have been dedicated to constructing high-quality explanation datasets. Techniques such as word overlap analysis \cite{peter}, LSH-based near-duplicate detection \cite{li2021extra} and a combination of manual and automatic reformulation on dialogue datasets \cite{guo2023towards} have been employed to this end.

Recently, the burgeoning field of LLMs has spurred research on LLM-based explainable recommendation, which still predominantly employs post-hoc approaches. For instance, \cite{chatrec} generates explanations in a zero-shot manner within a conversational scenario. \cite{preliminary} probes the innate reasoning capabilities of LLMs for recommendation tasks using in-context learning techniques. However, these approaches heavily rely on LLM's intrinsic reasoning capabilities, with the recommender system remaining unaware of the generated explanations, let alone assessing their quality. This raises concerns about their effectiveness and the accuracy of the produced justifications in reflecting the true reasoning behind recommendations.
This paper aims to address these limitations by proposing a novel approach that ensures coherent and reliable explanations directly integrated within the recommendation process.

\section{Preliminary}
\subsection{Problem Definition}
Let $\mathcal{U} = \{u_1, u_2, \ldots ,u_n\}$ denote a set of $n$ users and $\mathcal{I} = \{i_1, i_2, \ldots ,i_m\}$ a set of $m$ items. The user-item interaction data $\mathcal{D}$ is represented by a binary interaction matrix $\mathcal{R} \in \{0, 1\}^{n \times m}$, where $\mathcal{R}_{u, i}$ indicates whether user $u$ has interacted with item $i$. A value of 1 signifies explicit feedback (e.g., watching videos, clicking) and 0 otherwise. Each interaction is associated with a textual review $e_{u, i}$. The objective of explainable recommendation is to jointly predict future user interactions and generate explanations for these predictions. We formulate this as a probabilistic model:
\begin{align}
     P(\mathcal{Z}, \hat{y} \vert \mathcal{D})
\end{align}
where $\mathcal{Z}$ represents the set of explanations for all user-item pairs and $\hat{y}$ denotes the predicted interaction scores.

\subsection{Theoretical Basis of ExpCTR}
\label{theory}
To generate post-hoc explanations, we decompose the joint probability as follows:
\begin{align}
    P(\mathcal{Z}, \hat{y} \vert \mathcal{D}) = \underbrace{P(\hat{y} \vert \mathcal{D})}_{\text{CTR Model}} \cdot\underbrace{{P(\mathcal{Z} \vert \hat{y}, \mathcal{D})}}_{Generator}.
\end{align}
We first train a CTR model $f: \mathcal{U} \times \mathcal{I} \rightarrow \mathbb{R}$. This model learns latent representations $\mathbf{h}_{i,j}$ from user-item interaction and side information (e.g., user demographics, item features). The optimization process is formulated as:
\begin{align}
\min \sum_{(u,i) \in \mathcal{D}} \mathcal{L}_{CTR}(\hat{y}, y),
\end{align}
where $y$ denotes the ground truth interactions for user-item pairs. 
We define a generator $g: \mathcal{U} \times \mathcal{I} \times \mathbb{R} \rightarrow \mathcal{V}$ that explains why user $u$ might interact positively or negatively with item $i$. This model generates explanations conditioned on the predicted result $\hat{y}$. The generator is optimized as:
\begin{align}
     \min \sum_{(u, i) \in \mathcal{D}} \sum_{k=1}^{|\overline{e}_{u,i}|} -\log p(t_k \vert t_{<k}, \hat{y}),
\label{origin_ctr}
\end{align}
where $\overline{e}_{u,i}$ denotes the processed customer reviews used as explanation samples \cite{chen2019co, peter, li2021extra}.
Post-hoc methodologies exhibit a critical dependency on curated training datasets, which fundamentally shapes the conditional distribution $P(\mathcal{Z} \vert \hat{y}, \mathcal{D})$. Another limitation lies in that the generated explanations exert no influence on the CTR model, thereby failing to guarantee that the explanations faithfully reflect the underlying mechanisms of the CTR model. 
To address this, we integrate CTR prediction and explanation generation within a unified framework, leveraging LLMs, which can be mathematically expressed as:
\begin{align}
    P(\mathcal{Z}, \hat{y} \vert \mathcal{D}) = \underbrace{P(\mathcal{Z} \vert \mathcal{D})}_{\text{LLM}} 
    \cdot \underbrace{P(\hat{y} \vert \mathcal{Z}, \mathcal{D})}_{\text{CTR Model}}.
\end{align}
We employ a LLM to generate explanations, circumventing the need for constructing a high-quality explanation dataset – a laborious and costly task. The CTR model, $P(\hat{y} \mid \mathcal{Z}, \mathcal{D})$, depends on the generated explanations $\mathcal{Z}$, thus establishing a direct link between the explanations and their impact on CTR predictions. This approach represents a significant departure from traditional post-hoc methods, as the generated explanations are not simply after-the-fact rationalizations but integral components of the recommendation decision-making process.

Concretely, we adapt the CTR model to incorporate the generated explanations as features, denoted by $\overline{f}: \mathcal{U} \times \mathcal{I} \times \mathcal{V} \rightarrow \mathbb{R}$. The prediction is then computed as follows:
\begin{align}
\hat{y}= \overline{f}(\mathcal{R}, \mathcal{Z} \vert \Theta_{\text{CTR}}).
\label{new_ctr}
\end{align}

\begin{figure*}[t]
    \centering
    \includegraphics[width=16cm]{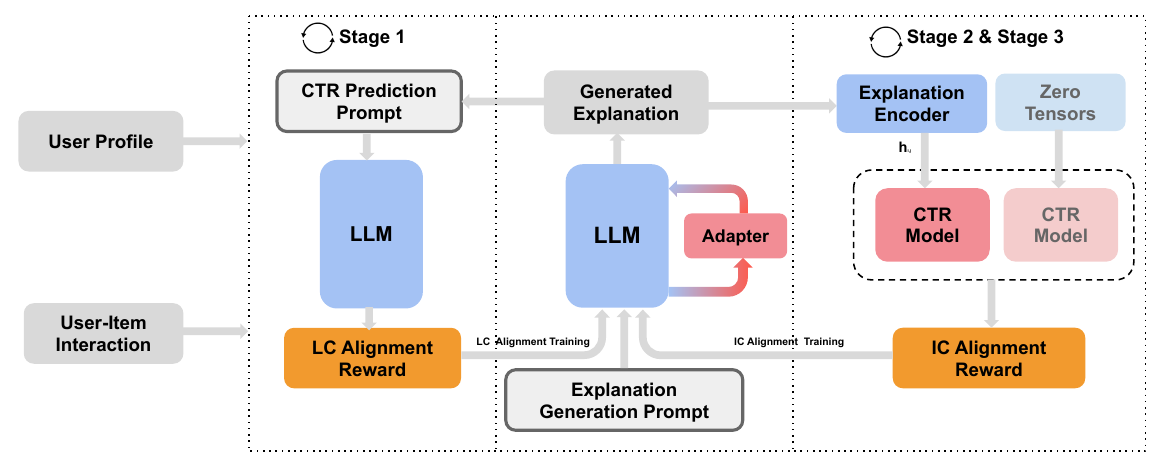}
    \caption{The overall architecture of ExpCTR.
    }
    \label{fig:enter-label}
\end{figure*}

\section{Methodology} 
Figure \ref{fig:enter-label} depicts the overall architecture of ExpCTR. It consists of three primary components: (1) Explanation Generation leverages a LLM to produce textual explanations for recommendations. (2) Reward Design utilizes CTR prediction processes to provide quality assessments for the generated explanations that serve as reward signals. (3) Training Paradigm introduces Lora lightweight fine-tuning techniques along with an iterative training process.

\subsection{Explanation Generation}
\label{sub:explanation generation}
Traditional recommendation systems rely on implicit representations of users and items and suffer from a lack of interpretability \cite{ferrari2019we}. However, recent advancements in LLMs have demonstrated extensive world knowledge and advanced reasoning capabilities \cite{reasoning1, reasoning2, reasoning3}. These capabilities offer a promising avenue for human-interpretable explanation generation. To harness this potential, we design a prompt template to guide the LLM to generate effective explanations. The prompt leverages user historical interaction data and frames the LLM as a helpful recommendation assistant:
\begin{tcolorbox}[title = {Prompt Template for Explanation Generation}]
You are a helpful online recommendation assistant with access to a vast database of books and reader reviews. A customer has provided you with information about their reading preferences: \\
    The liked books: <item\_1\_1>$\ldots$ \\
    The disliked books: <item\_2\_1>$\ldots$ \\
Considering the preference of the customer, predict how the customer will consider \textit{<Target Item>} and give a reason. The answer should be within one sentence.
\end{tcolorbox}
The prompt template, originally developed for a book recommendation task \cite{yu2024ra,yu2024break}, can be easily adapted to different recommendation scenarios with minor adjustments. Items (e.g., <item\_1\_1>$\ldots$) are represented by their titles, while users are characterized by the titles of items they have interacted with. To further refine user profiles, we categorize these historical items into liked and disliked categories, using interaction signals such as ratings as indicators. A threshold-based function is employed to classify each item in a user’s interaction sequence. Items rated above the threshold are considered 'liked', while those below are deemed 'disliked'. The threshold is a hyperparameter adjusted based on dataset characteristics.

This structured prompt empowers the LLM to effectively utilize its knowledge base to infer more nuanced user preferences from the liked and disliked items and generate rationales for why a user might like or dislike a particular target item. By applying this methodology to all user-item interaction pairs $\mathcal{D}$, we obtain a set of explanations $\mathcal{Z}$, where each explanation $\mathcal{Z}_{u,i} \in \mathcal{Z}$ corresponds to a specific user-item pair.

\subsection{Reward Design}
We optimize the LLM through quality assessment of the generated explanations. Inspired by InstructGPT \cite{instructgpt}, we leverage a reinforcement learning paradigm to achieve this objective with a well-designed reward function that incentivizes the LLM to generate informative explanations for CTR prediction and accurately represent the underlying user motivations behind their interactions. 

\subsubsection{Proximal Policy Optimization}
Following \cite{instructgpt}, we adopt the proximal policy optimization (PPO) \cite{ppo} algorithm for the reinforcement learning process. Given a prompt and response (explanation), the LLM produces a reward determined by a reward function, concluding the episode. The objective function for PPO training is formulated as:
\begin{equation}
\begin{aligned}
    \text{objective}_{\phi} = & E_{(x,\mathcal{Z}) \sim D_{\pi_\phi^{\text{RL}}}} \left[R(\mathcal{Z}) - \beta \log \frac{\pi_\phi^{\text{RL}}(\mathcal{Z}|x)}{\pi^{\text{init}}(\mathcal{Z}|x)}\right],
\end{aligned}
\end{equation}
where $x$ denotes the prompt template for explanation generation, as detailed in Section \ref{sub:explanation generation}. $\pi^{\text{init}}$ is the initial LLM and $\pi_\phi^{\text{RL}}$ represents the fine-tuned explanation generation language model to be optimized.$\beta$ is the KL penalty and $R(\mathcal{Z})$ is our reward function.

The core concept behind our reward function design is to leverage real-world CTR prediction feedback to enhance explanation quality. We aim to incentivize explanations that accurately reflect user intent and preferences. Crucially, these explanations must also align with the underlying mechanics and predicted outcomes of the CTR models employed by the recommender system. To accomplish this, we decompose the reward function into two key components: Explanation and LLM-CTR Alignment, and Explanation and ID-CTR Alignment. These components will be elaborated in the following sections.

\subsubsection{Explanation and LLM-CTR Alignment (LC Alignment)}
This component evaluates the effectiveness of the LLM's generated explanation towards accurately inferring the intended user behavior. A high LC alignment reward signifies that the explanation successfully conveys the underlying factors influencing user interaction. We operationalize LC alignment reward by leveraging recent advancements in CTR prediction with LLMs. We frame the task as a binary classification problem, where the LLM predicts whether a user will like a given item (e.g., book) based on their rationales (the generated explanation by LLM in Section \ref{sub:explanation generation}).

Specifically, we design a prompt template to guide the LLM towards predicting CTR. This template provides context for the LLM, including the user's thoughts about the item and a binary response option ("Yes" or "No") indicating their decision:
\begin{tcolorbox}[title = {Prompt Template for CTR Prediction}]
Given how the user thinks about a book, identify whether the user will like the target book by answering "Yes." or "No.". \\
The user thought: \textit{<reason>} \\
Whether the user will like the target book {target}:
\end{tcolorbox}
In this template, \textit{<reason>} is replaced with the explanation $\mathcal{Z}$ from Section \ref{sub:explanation generation}. Formally, we define the predicted CTR score for a user-item pair $(u, i)$ as:
\begin{align}
    s^u_{u, i} = \frac{exp({p(t_0= \mathcal{V}_{pos}) / T})}{exp({p(t_0= \mathcal{V}_{pos}) / T}) + exp({p(t_0= \mathcal{V}_{neg}) / T})},
\end{align}
where $p(t_0= \mathcal{V}_{pos})$ is the probability of the first generated token $t_0$ by LLM that equals $\mathcal{V}_{pos}$.
$T$ is the temperature for softmax function and $\mathcal{V}_{pos} = \{"Yes"\}, \mathcal{V}_{neg} = \{"No"\}$.

A closer alignment between the CTR prediction and the ground-truth label indicates a more precise explanation, demonstrating the ability to capture the actual factors influencing user's decisions. We formulate the LC alignment reward as:
\begin{align}
    R_{LC}(\mathcal{Z}_{ui}) = 1 - \vert y_{u, i} - s^u_{u, i} \vert.
\label{correct reward}
\end{align}
However, directly using this reward function might lead to unstable gradients due to potential variations in reward scales across different batches \cite{zheng2023secrets}. We introduce a normalization and clipping procedure to ensure that reward values are appropriately scaled and bounded:
\begin{align}
    R_{LC}^{norm}(\mathcal{Z}_{ui}) = \text{clip}\left(\frac{R_{LC}(\mathcal{Z}_{ui}) - \text{mean}\left({R_{LC}}(\mathcal{Z}_{ui})\right)}{\text{std}(R_{LC}(\mathcal{Z}_{ui}))}, \delta \right),
\label{reward_norm}
\end{align}
where $\text{mean}\left({R_{LC}}(\mathcal{Z}_{ui})\right)$ and $\text{std}(R_{LC}(\mathcal{Z}_{ui}))$ denotes the mean and standard deviation of the rewards across a batch. The clip function constrains the normalized reward within a predefined bound $\delta$.

\subsubsection{Explanation and ID-CTR Alignment (IC Alignment)}

The congruence between generated explanations and the CTR model is quantitatively assessed by evaluating their contribution to CTR predictions. A positive reward value potentially signifies that the explanation provides substantial insights into the underlying patterns driving CTR. To rigorously evaluate this alignment, we integrate the generated explanations directly into the existing CTR prediction architecture. This integration serves a dual purpose: evaluating explanatory quality and potentially enhancing predictive accuracy by leveraging latent information within the explanations.

Our approach first obtains a dense textual representation for each generated explanation $\mathcal{Z}_{u, i}$ by employing 
a pre-trained language model (PLM), $f_{encoder}: \mathcal{V} \rightarrow \mathbb{R}^{d}$, to map the explanation text into a unified semantic space. This enables the capture of underlying meaning and relationships within the explanation.  $f_{encoder}$ can be any frozen pre-trained language model, such as BERT \cite{devlin2018bert}, BGE \cite{bge_embedding} and we derive the dense representation for the explanation as follows:
\begin{align}
    \mathbf{z}_{u, i} = \text{MeanPooling}(f_{encoder}(\mathcal{Z}_{u, i})),
\end{align}
where $\mathbf{z}_{u, i}$ denotes the mean pooling hidden representations from the last layer in PLM.

Subsequently, we integrate these textual representations with an original ID-based CTR model architecture. This integration facilitates the learning of a joint representation that combines user and item information with the insights provided by the explanation. We propose a simple yet effective concatenation operation to achieve this integration. Specifically, the textual representation $\mathbf{z}_{u, i}$ is concatenated with the hidden representation $\mathbf{h}_{u, i}$ (defined in Section \ref{theory}) and fed into the existing CTR model to predict the CTR score as follows:
\begin{align}
    s^r_{u,i} = \overline{f}(\text{Concate}(\mathbf{h}_{u, i}, \mathbf{z}_{u, i})),
\end{align}
where $\overline{f}$ (as in Equation \ref{new_ctr}) can be any original ID-based model architecture, such as DeepFM \cite{lian2018xdeepfm}. To evaluate the impact of semantic representations of LLM's explanations, we compare the performance of the CTR model with and without these explanations and quantify the differences in CTR predictions. A notable performance improvement when explanations are incorporated indicates that the introduced semantic features contribute positively. This implies a better-aligned explanation, justifying a higher reward. This evaluation is formalized as follows:
\begin{align}
    R_{IC}(\mathcal{Z}_{u, i}) = 1 - \vert y_{u, i} - s^r_{u,i} \vert + \vert s^r_{u,i} - \Tilde{s}^r_{u,i} \vert,
\label{useful reward}
\end{align}
where $\Tilde{s}^r_{u,i}$ indicates the CTR prediction score obtained without using explanations as input features, by setting the representations $\mathbf{z}_{u, i}$ to zero vectors. The IC alignment reward is normalized and clipped, as in Equation \ref{reward_norm}, resulting in $R^{norm}_{IC}(\mathcal{Z}_{u, i})$.

By synergizing the effects of these two reward components during the LLM training phase, we aim to ensure that the generated explanations not only accurately capture user rationales behind their behavior but also contribute meaningfully to the performance of CTR models. This approach promotes explanations that are both faithful and informative, ultimately leading to a more robust and interpretable recommendation system.

\subsection{Training Paradigm}

\subsubsection{Light Weight Tuning}
To mitigate the computational training burden associated with three independent LLMs - the initial LLM $\pi^{init}$, explanation generator $\pi_\phi^{\text{RL}}$ and the LC alignment reward model, we adopt a lightweight tuning approach.
Recent findings \cite{houlsby2019parameter, li2021prefix, hu2021lora} suggest that LLMs can be effectively compressed without significant performance degradation, owing to the inherently lower-dimensional nature of the information they encode.
Leveraging this insight, we employ Low-Rank Adapters (Lora) \cite{hu2021lora} to optimize our training process which introduces trainable low-rank matrices into each transformer layer, allowing for efficient parameterization while preserving model performance.
Specifically, we employ a base LLM as both the initial model and the frozen LC alignment reward model. The explanation generator is instantiated as the base LLM with Lora.
Crucially, this strategy drastically reduces the parameters. By consolidating computations into a single LLM with a minimal number of trainable parameters in Lora, we achieve substantial computational efficiency without compromising model quality.

\subsubsection{Iterative Training}
This section outlines the iterative training methodology employed to optimize the explanation generation model, $\pi_\phi^{\text{RL}}$, considering both LC alignment and IC alignment rewards. Our approach involves a three-stage iterative training process, alternating between component-specific optimization phases.

\noindent\textbf{Stage 1. LC Alignment.} We commence by utilizing a frozen language model $\pi^{init}$ to compute LC rewards. The explanation generation model $\pi_\phi^{\text{RL}}$ is then optimized using the LC alignment reward:
\begin{equation}
R({\mathcal{Z}}) = R_{LC}^{norm}(\mathcal{Z}_{ui}).
\end{equation}
This phase establishes a foundational understanding of "correct" explanations, aligning the model with user preferences and intentions, and producing factually sound explanations. 
This stage persists for a predetermined number of iterations, during which we continuously accumulate fresh explanations for each user-item pair.

\noindent\textbf{Stage 2. CTR Model Training with Textual Features.} Following Stage 1, we accumulate a corpus of generated explanations. These explanations are integrated as textual features with the original ID dataset for training the CTR model $\overline{f}$:
\begin{align}
    \min \sum_{(u,i, \mathcal{Z}_{u, i}) \in \{\mathcal{D}, \mathcal{Z}\}} \mathcal{L}_{CTR}(s^r_{u,i}, y_{u, i}).
\end{align}

\noindent\textbf{Stage 3. IC Alignment.} In the final stage, we further refine the explanation generation model by incorporating the IC alignment reward:
\begin{align}
    R({\mathcal{Z}}) = R_{IC}^{norm}(\mathcal{Z}_{ui}).
\end{align}
Stages 2 and 3 are then repeated for a predefined number of iterations, allowing for continuous model refinement and performance improvement.
Upon the completion of the training process, we obtain a robustly trained explanation generation model $\pi_\phi^{\text{RL}}$ and a CTR model $\overline{f}$ that effectively leverages textual features for prediction.
This unified training approach prioritizes that generated explanations are informative and likely to resonate with both users and the CTR model, avoiding the pitfall of producing generic or uninformative content.

\begingroup
\setlength{\tabcolsep}{5pt} 
\renewcommand{\arraystretch}{0.95} 
\begin{table*}[!t]
\centering
\caption{Overall performance of different recommendation approaches on three benchmark datasets. The best results within the baseline methods and all methods are highlighted with \underline{underlined} and \textbf{boldface}.
ExpCTR-LLM and ExpCTR-Aug denote the CTR performance of the LC alignment and IC alignment reward model, respectively.
} \label{table:main}
\begin{tabular}{l | c c c c | c c c c | c c c c}
\toprule[1.5pt]
\multirow{2}{*}{Models} & \multicolumn{4}{c|}{BookCrossing} & \multicolumn{4}{c|}{ML-20M} & \multicolumn{4}{c}{Amazon Books} \\ \cmidrule{2-13}
 & AUC  & LogLoss  & MAE & RMSE & AUC  & LogLoss  & MAE & RMSE & AUC  & LogLoss  & MAE & RMSE  \\ \midrule
FM & 0.5176 & 0.6950 & 0.4991 & 0.5009 & 0.6231 & 0.6625 & 0.4702 & 0.4849 & 0.5667 & 0.6876 & 0.4950 & 0.4972 \\
DeepFM & 0.5222 & 0.6924 & 0.4987 & 0.4996 & 0.6212 & 0.6708 & 0.4762 & 0.4891 & 0.5639 & 0.6873 & 0.4956 & 0.4971 \\
AutoInt & 0.5176 & \underline{0.6922} & 0.4990 & \underline{0.4995} & 0.6324 & 0.6636 & 0.4618 & 0.4851 & 0.5678 & 0.6868 & 0.4957 & 0.4968 \\
PNN & 0.5257 & 0.6926 & 0.4994 & 0.4997 & 0.6363 & 0.6786 & 0.4575 & 0.4917 & 0.5733 & \underline{0.6860} & 0.4938 & \underline{0.4964} \\
xDeepFM & 0.5124 & 0.6980 & 0.4999 & 0.5024 & 0.6440 & 0.6628 & 0.4645 & 0.4851 & 0.5640 & 0.6912 & 0.4905 & 0.4987 \\
FiGNN & 0.5211 & 0.6922 & 0.4989 & 0.4997 & 0.6416 & 0.6762 & 0.4751 & 0.4917 & 0.5673 & 0.6893 & 0.4919 & 0.4980 \\
DCN & 0.5198 & 0.6954 & 0.4987 & 0.5011 & 0.6250 & 0.6891 & 0.4465 & 0.4942 & 0.5487 & 0.7120 & 0.4906 & 0.5082 \\
DCNV2 & 0.5195 & 0.7188 & 0.4943 & 0.5113 & 0.6134 & 0.6800 & 0.4806 & 0.4935 & 0.5433 & 0.7230 & 0.4907 & 0.5132 \\
DIN & 0.5132 & 0.7660 & 0.4931 & 0.5308 & 0.6033 & 0.7540 & \underline{0.4449} & 0.5134 & 0.5120 & 0.8968 & 0.4945 & 0.5600 \\
DIEN & 0.5231 & 0.7353 & 0.5030 & 0.5200 & 0.6074 & 0.6798 & 0.4654 & 0.4925 & 0.5096 & 0.9111 & 0.4962 & 0.5572 \\
\midrule
CASER & 0.5208 & 0.6931 & 0.4997 & 0.5000 & 0.6407 & 0.6836 & 0.4949 & 0.4952 & 0.5205 & 0.6919 & 0.4991 & 0.4994 \\
GRU4Rec & 0.5356 & 1.3650 & 0.4911 & 0.5664 & 0.6403 & 0.6902 & 0.4985 & 0.4985 & 0.5283 & 0.6930 & 0.4999 & 0.4999 \\
SASRec & 0.5322 & 1.1634 & 0.4864 & 0.5771 & 0.6197 & 0.6695 & 0.4817 & 0.4882 & 0.5181 & 0.7145 & 0.4932 & 0.5072 \\
BERT4Rec & 0.5136 & 1.0717 & 0.5017 & 0.6075 & 0.5866 & 0.6789 & 0.4885 & 0.4929 & 0.5298 & 0.7547 & 0.4923 & 0.5130 \\
\midrule
TALLRec & 0.5389 & 0.6929 & 0.4969 & 0.5005 & \underline{0.6660} & \underline{0.6541} & 0.4726 & \underline{0.4804} & 0.5744 & 0.6868 & 0.4955 & 0.4968 \\
ICL & \underline{0.5663} & 0.7328 & \underline{0.4829} & 0.5174 & 0.6320 & 0.6754 & 0.4556 & 0.4908 & \underline{0.5930} & 0.7246 & \underline{0.4715} & 0.5133 \\
\midrule
ExpCTR-LLM & 0.6042 & 0.6943 & 0.4734 & 0.4999 & 0.6707 & 0.6428 & 0.4523 & 0.4749 & 0.6290 & 0.6831 & 0.4720 & 0.4946 \\
ExpCTR-Aug & \textbf{0.6173} & \textbf{0.6715} & \textbf{0.4800} & \textbf{0.4891} & \textbf{0.6951} & \textbf{0.6389} & \textbf{0.4210} & \textbf{0.4710} & \textbf{0.6641} & \textbf{0.6493} & \textbf{0.4557} & \textbf{0.4783} \\
\bottomrule[1.5pt]
\end{tabular}
\label{exp_ctr}
\end{table*}

\section{Experiment}
In this section, we detail the experimental setup to evaluate the performance of ExpCTR. We aim to address the following research questions through a series of rigorous experiments and analyses:
\begin{itemize}[leftmargin=*]
    \item RQ1: How does ExpCTR compare to existing state-of-the-art approaches in terms of generating explanations for recommendation decisions and improving CTR prediction?
    \item RQ2: How effective is the integration of the PPO algorithm in ExpCTR?
    \item RQ3: How does the quality of the explanations produced by our framework measure up?
\end{itemize}

\subsection{Experimental Setting}

\subsubsection{Datasets}
To comprehensively evaluate the effectiveness and generalizability of our proposed framework, we leverage three publicly available, large-scale datasets: BookCrossing \footnote{https://www.kaggle.com/datasets/somnambwl/bookcrossing-dataset}, MovieLens-20M \footnote{https://grouplens.org/datasets/movielens/20m/}, and Amazon Books \footnote{https://jmcauley.ucsd.edu/data/amazon/}. Following \cite{bao2023tallrec},
we employ a stratified random sampling approach for each user within each dataset. Specifically, we randomly select one item a user interacted with as the target item for prediction. The remaining interacted items, up to a maximum of 10 items chronologically preceding the target item, are considered the user's historical interactions. Then, we partition the constructed data samples into training, validation, and testing sets with a ratio of 8:1:1. For datasets containing rating scores, we binarize the ratings using a threshold where ratings above the threshold are considered positive interactions (items the user liked), while ratings below are considered negative interactions. Specifically, the threshold for the ML-20M and Amazon Books datasets is 4, and 5 for the BookCrossing dataset \cite{song2019autoint, dien}. Finally, Amazon Books and ML-20M comprise 16,000/2,000/2,000 while BookCrossing comprises 32,000/4,000/4,000 data samples.

\subsubsection{Compared Methods}
For CTR evaluations of ExpCTR, we leverage two distinct scoring mechanisms: LLM scores derived from the LC alignment module, designated as "ExpCTR-LLM", and CTR scores with explanations as textual features from the IC alignment module, referred to as "ExpCTR-Aug". ExpCTR-LLM reflects the effectiveness of the generated explanations in capturing and articulating user preferences and rationales for future interactions, which results in better outcomes under an LLM scorer.
Conversely, a superior ExpCTR-Aug score suggests that the explanation aligns well with the internal workings of ID-based CTR models and provides supplementary information that enhances performance. This dual evaluation approach provides an indirect yet effective method for assessing explanation quality.
We compare ExpCTR against diverse established baseline models, encompassing both ID-based and LLM-based recommendation methods:
\begin{itemize}[leftmargin=*]
    \item ID-based methods: Factorization Machines (FM) \cite{fm} captures pairwise feature interactions for recommendation tasks. Deep learning models, including DSSM \cite{DSSM}, DeepFM \cite{lian2018xdeepfm}, AutoInt \cite{song2019autoint}, PNN \cite{pnn}, Fi-GNN \cite{fignn}, DCN \cite{dcn}, DCNV2 \cite{wang2021dcnv2}, utilize multi-layer perceptrons, self-attention mechanisms, and graph neural networks to effectively capture both low-order and high-order feature interactions to enhancing recommendation accuracy. DIN \cite{zhou2018din} and DIEN \cite{dien} leverage attention mechanisms to extract user dynamic interests from their historical behavior sequences.  Caser \cite{tang2018caser}, GRU4Rec \cite{gru4rec}, SASRec \cite{sasrec} and BERT4Rec \cite{sun2019bert4rec} are sequential-based recommendation models that employ Convolutional Neural Networks (CNNs), Gated Recurrent Units (GRUs), and transformer-encoder architectures for robust user behavior modeling, respectively, leading to more accurate recommendations.
    \item LLM-based methods: In-Context Learning (ICL) for Recommendation \cite{icl} leverages an LLM for recommendations by directly posing queries to the LLM. TALLRec \cite{bao2023tallrec} adapts LLMs to recommendation scenarios through instruction tuning.
\end{itemize}

\subsubsection{Metrics}
To assess the effectiveness of ExpCTR, we utilize multiple regular CTR prediction metrics \cite{lian2018xdeepfm, zhou2018din}. Specifically, we evaluate performance using the Area Under the ROC Curve (AUC), binary cross-entropy loss (Log Loss), Mean Squared Error (MSE), and Root Mean Squared Error (RMSE).

\subsubsection{Implementation Details}
In our experimental setup, we employ LLaMA-3-7b as the foundational model for both explanation generation and LC alignment reward computation. For explanations encoding, we employ BGE-small \cite{bge_embedding}. The IC alignment reward is built upon the DeepFM and implemented through the open-source project Recbole \cite{recbole[1.0]}. Our optimization incorporates a learning rate of $1\times10^{-5}$, with a KL penalty of 0.05. The reward clip threshold $\delta$ is set to 1.0. The iterative training paradigm consists of two epochs per iteration. 
For TALLRec, we leverage the entire training dataset and apply a learning rate of $1\times10^{-4}$. All experiments are conducted on a single machine equipped with NVIDIA A800 GPUs.

\subsection{Performance Comparison (RQ1)}
Table \ref{exp_ctr} presents a comparative analysis of our proposed method with existing ID-based CTR methods and LLM-based methods. The results yield several noteworthy observations:

\begin{itemize}[leftmargin=*]
    \item Baseline models such as ICL and TALLRec demonstrate strong performance across all datasets, particularly when compared to ID-based methods. This suggests that the LLMs possess a robust foundational capability for reasoning and comprehension. Nevertheless, ExpCTR-LLM consistently surpasses these two LLM-based CTR models on all metrics and datasets. This empirical evidence indicates that our generated explanations accurately reflect and describe user behavior patterns, leading to significant performance improvements over ICL, which uses the same frozen LLM scorer, and TALLRec, which is finetuned directly under the CTR prediction task.
    These findings highlight ExpCTR's capability to leverage the intrinsic reasoning capabilities of LLMs effectively. Additionally, the consistent performance gains underscore the efficacy of our proposed framework, with the integration of reinforcement learning and the LC alignment reward function further extending and motivating the potential of LLMs in recommendation scenarios.

    \item ExpCTR-Aug emerges as a substantial advancement over ExpCTR-LLM, demonstrating superior performance across all evaluated datasets. This result highlights the pivotal role of the IC alignment reward in augmenting model efficacy. The explanations generated by ExpCTR-Aug offer profound insights into ID-based CTR models, leading to considerable performance improvements compared to the DeepFM baseline, with observed gains of 18.2\%, 11.9\%, and 17.8\% in AUC across the respective datasets. 
    These results underscore the dual advantages of ExpCTR that it not only enhances the interpretability of the recommendation system but also delivers substantial improvements in recommender system accuracy.
\end{itemize}


\subsection{In-depth Analysis of PPO (RQ2)}
\subsubsection{PPO Reward Analysis}

\begin{figure}[!t]
  \centering
  \subfigure[\scriptsize LC Alignment Reward on ML-20M]{
    \includegraphics[width=0.22\textwidth]{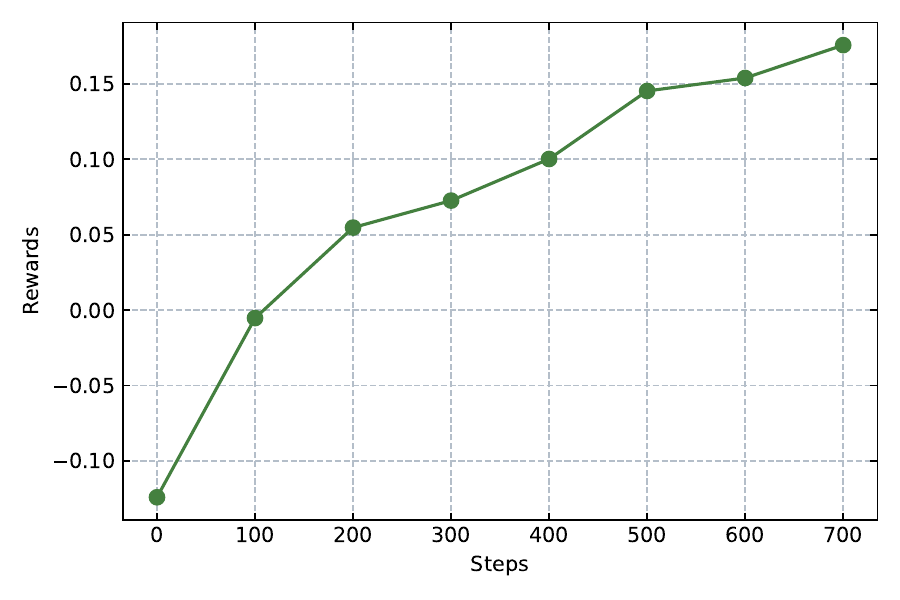}
  }
  \subfigure[\scriptsize IC Alignment Reward on ML-20M]{
    \includegraphics[width=0.22\textwidth]{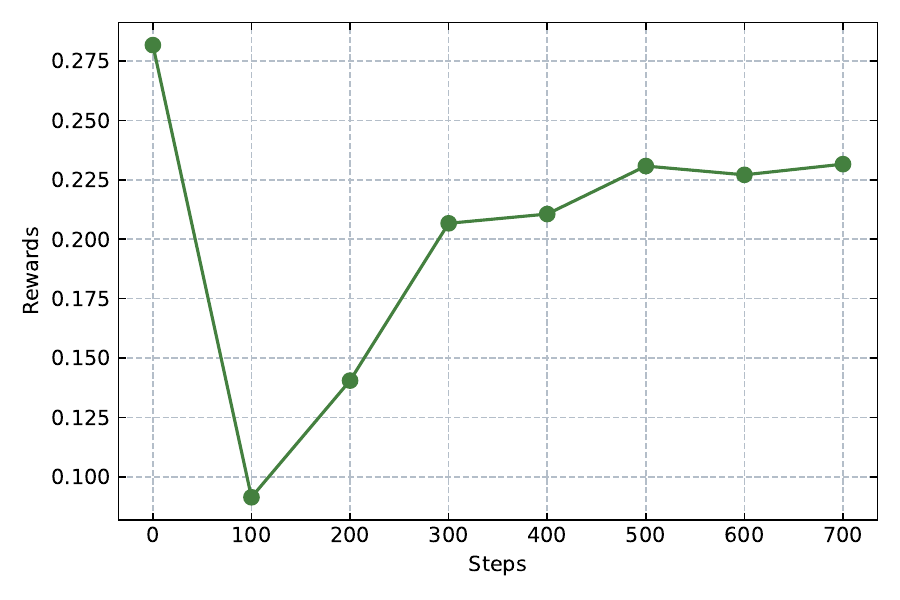}
  }
  \subfigure[\scriptsize LC Alignment Reward on BookCrossing]{
    \includegraphics[width=0.22\textwidth]{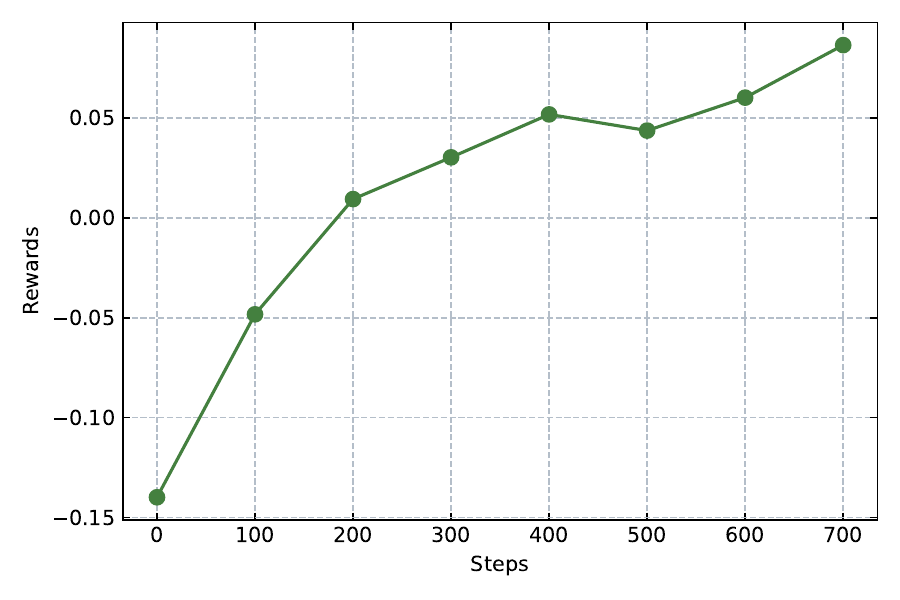}
  }
  \subfigure[\scriptsize IC Alignment Reward on BookCrossing]{
    \includegraphics[width=0.22\textwidth]{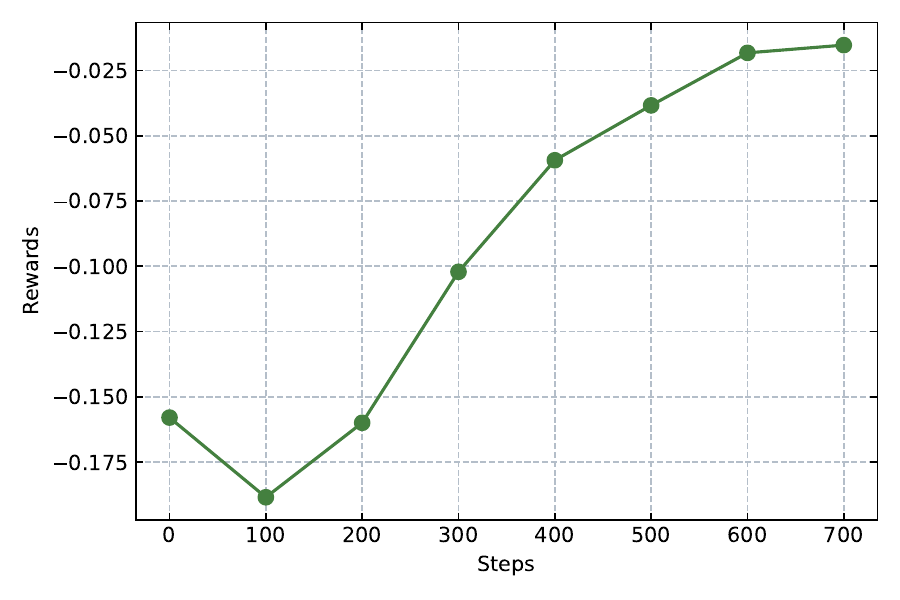}
  }
\caption{Trends of LC alignment and IC alignment rewards across iterative training on ML-20M and BookCrossing datasets.}
\label{rewards_curve}
\end{figure}

To investigate the efficacy and learning dynamics of the training paradigm in ExpCTR, we conduct a comprehensive analysis of the reward trajectories during the iterative training process. Specifically, we examined the evolution of LC alignment and IC alignment rewards on ML-20M and BookCrossing datasets. The results of this analysis are presented in Figure \ref{rewards_curve}.

Our analysis reveals a consistent upward trend in LC alignment rewards across both datasets, which stabilizes during the final stages of training and coincides with the performance enhancement of ExpCTR-LLM. Notably, the ML-20M dataset exhibits a more pronounced increase, ultimately reaching a higher plateau. This observation aligns with the superior performance obtained on the ML-20M dataset (9.1\% improvement in AUC over ICL). These observations suggest that our LC alignment reward mechanism effectively steers LLMs towards generating explanations that are increasingly congruent with user behavior patterns and exhibit a strong correlation with subsequent user interactions.
In contrast, the IC alignment stage is characterized by a more fluctuating curve across both datasets, with the alignment stabilizing more rapidly compared to LC alignment. This corresponds to the relatively modest improvement observed in ExpLLM-Aug over ExpCTR-LLM. The BookCrossing dataset experiences a slight decline followed by steady growth, reflecting the refinement process of IC alignment for LLM-based explanation generation.
This empirical evidence underscores the effectiveness of our training paradigm in fostering the development of a more user-centric and contextually relevant recommendation system.

\subsubsection{Hyperparameter Sensitivity Analysis}

\begin{figure}[!t]
  \centering
  \subfigure[\scriptsize KL Penalty $\beta$]{
    \includegraphics[width=0.22\textwidth]{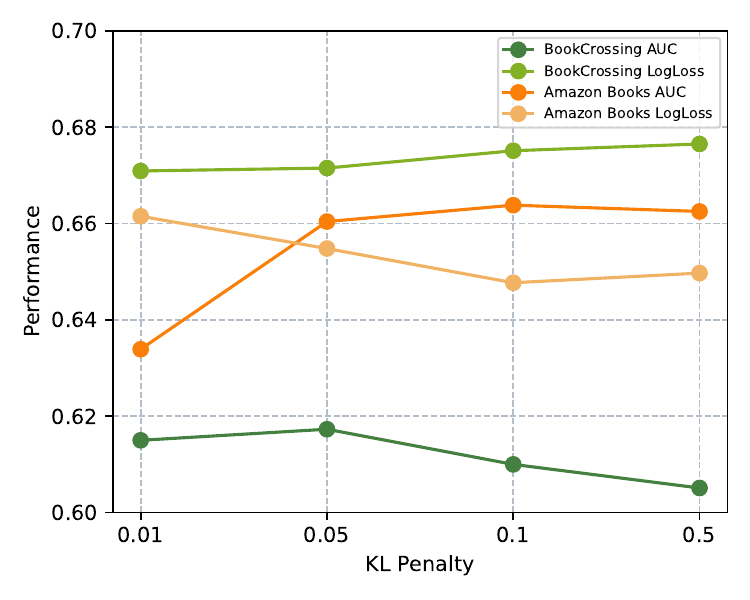}
  }
  \subfigure[\scriptsize Reward Normalization Bound $\delta$]{
    \includegraphics[width=0.22\textwidth]{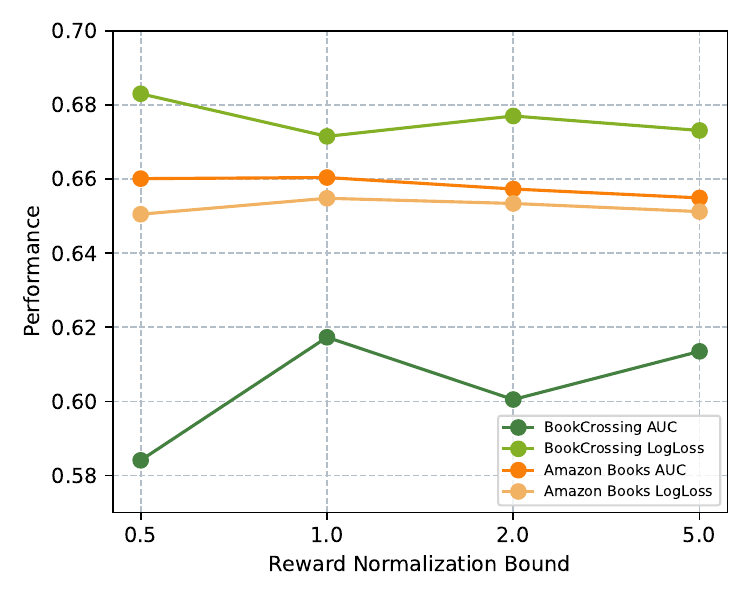}
    }
\caption{Hyperparameter sensitivity of ExpCTR in KL penalty $\beta$ and reward normalization bound $\delta$.}
\label{hyper}
\end{figure}

We assess the sensitivity of hyperparameters in PPO training, specifically focusing on the KL penalty $\beta$ and the reward normalization bound $\delta$, both of which are crucial for effective PPO training \cite{ppo}. Figure \ref{hyper} shows the performance variations across different hyperparameter settings, with $\beta$ ranging from [0.01, 0.05, 0.1, 0.5] and $\delta$ ranging from [0.5, 1.0, 2.0, 5.0], evaluated on the BookCrossing and Amazon Books datasets.
For the KL penalty $\beta$, its effect on ExpCTR's performance is notable across both datasets. Specifically, extreme values of $\beta$—either too high or too low—detract from the model's capabilities, with a setting of 0.05 typically yielding the most competitive results. In contrast, the reward normalization bound $\delta$ shows significant performance variability on the BookCrossing dataset, while remaining stable on the Amazon Books dataset and we choose $\delta=1.0$ for both datasets.

\begin{table}[t]
\centering
\scriptsize
\setlength{\tabcolsep}{2pt} 
\caption{Explanation generated by ICL and ExpCTR on Amazon Books dataset.}
\renewcommand{\arraystretch}{1.1} 
\begin{tabular}{p{1cm}|p{7cm}}
\toprule[1.5pt]
Review & \hl{The right book}, but I would have had to return it for the code that comes with it. My class had already started. All materials should have been included. \\ \midrule
ICL & Based on the customer's interest in books about crystals, gemstones, and nursery rhymes, I predict that they will have a \hl{neutral or indifferent opinion} about "Real Research: Conducting and Evaluating Research in the Social Sciences" as this book's topics and focus on rigorous research methods and scientific inquiry may not align with their preferred themes.
 \\ 
\midrule
Ours & Based on the customer's preferences for books like "Crystal Lore" and "Legends \& My Thrills: The Fascinating History of the World's Most Powerful Gems and Stones", I predict that they will \hl{likely enjoy} "Real Research: Conducting and Evaluating Research in the Social Sciences" and consider it a strong fit for their preferences because it is a non-fiction book that offers a sense of discovery and exploration, similar to the themes of \hl{history and mythology} in their preferred books, and involves a sense of mystery and discovery that is also present in the nursery rhyme book, "Jack and Jill", which they also felt drawn.  \\
\end{tabular}

\begin{tabular}{p{1cm}|p{7cm}}
\toprule[1.5pt]
Review &  $\ldots$I am happy to say that the \hl{sharp wit, business savvy}, and zeal that I experienced in my conversations with Gary come across in Crush It! $\ldots$ Clearly, anyone new to the world of social media will \hl{find this book informative, instructive, and easy to read}$\ldots$ I suppose I am a great example of the latter, $\ldots$ regularly speak to \hl{business and professional groups}, and use these techniques every day and I have a page of notes that I made while reading this book. \\ \midrule
ICL & I predict that this customer will \hl{likely dislike} "Crush It!: Why NOW Is the Time to Cash In on Your Passion" because they have shown a preference for practical, non-fiction books focused on leadership and strategy (e.g. "Surviving to Thriving" and "The Expert's Edge"), and "Crush It!" is a self-help book with a more entrepreneurial and inspirational tone that may not align with their reading interests. \\ 
\midrule
Ours & Based on the customer's preferences, I predict that they will \hl{likely enjoy} "Crush It!: Why NOW Is the Time to Cash In on Your Passion" and consider it a book that they will like, as it shares a similar tone and style of writing with the books they have liked, such as "Glissando" and "The Art of Woo", and will likely appeal to \hl{their interest in topics such as business, self-improvement, and marketing}, which is also present in the books they have liked, such as "Predictably Irrational" and "Made to Stick". \\
\bottomrule[1.5pt]
\end{tabular}
\label{data}
\end{table}

\subsection{Case Study (RQ3)}
To elucidate the efficacy of ExpCTR in generating improved explanations, we present a comparative analysis of explanations produced by ICL and our proposed approach. Table \ref{data} illustrates representative examples, accompanied by actual user reviews to provide real-world context for our analysis.

In the first case, we observe that the user's attitude towards the targeted book is fundamentally positive. The user's comment, \textit{"The right book"}, indicates approval, while the phrase \textit{"class have already started"} represents an extraneous factor beyond the scope of the recommender system. This positive sentiment aligns with the high CTR prediction of 0.8843. ICL gives a negative attitude (\textit{"neutral or indifferent opinion"}). However, ExpCTR successfully captures this alignment towards the CTR model, generating a recommendation explanation that accurately reflects the user's probable affinity for the book (\textit{"similar to the themes of history and mythology in their preferred books"}).
The second case demonstrates a more nuanced improvement. The ICL approach erroneously infers a negative attitude (\textit{"likely dislike"}), contradicting the user's actual 5.0 rating. In contrast, ExpCTR, enhanced by LC alignment reward training, correctly identifies the positive interaction potential (\textit{"likely enjoy"}). Furthermore, the explanation generated by our model corresponds closely to the user's actual thoughts, accurately identifying the book's themes of \textit{"business, self-improvement, marketing"}.

\section{Conclusion}
In the paper, we present ExpCTR to address the limitations of current post-hoc explainable recommendation methods. By integrating LLM-based explanation generation into the CTR prediction process, ExpCTR eliminates the need for extensive data preparation and mitigates reliability issues. Our approach leverages reinforcement learning to align LLM reasoning with both user preferences and the recommender system's internal workings. 
We believe that ExpCTR represents a significant step forward in the field of explainable recommendations and opens up new avenues for future research.

\bibliographystyle{ACM-Reference-Format}
\bibliography{sample-base}


\end{document}